\batchmode
\makeatletter
\def\input@path{{/Users/axelaraneda/Desktop/Research/Hurst/}}
\makeatother
\documentclass[english]{article}
\usepackage{lmodern}
\usepackage[T1]{fontenc}
\usepackage[latin9]{inputenc}
\usepackage[a4paper]{geometry}
\usepackage{units}
\usepackage{amsmath}
\usepackage{amsthm}
\usepackage{amssymb}
\usepackage{graphicx}
\usepackage[numbers,numbers,sort&compress]{natbib}

\makeatletter

\usepackage{tikz}
\usetikzlibrary{calc}

\newcommand{\lyxaddress}[1]{
	\par {\raggedright #1
	\vspace{1.4em}
	\noindent\par}
}

\usepackage{hyperref}
\date{}

\usepackage{todonotes}
\usepackage{bbm}
\usepackage{fontawesome5}
\usepackage{orcidlink}
\usepackage{doi}

\makeatother

\usepackage{babel}
\begin{document}
\title{\textbf{A multifractional option pricing formula}}
\author{Axel A.~Araneda\orcidlink{0000-0003-4436-7974}\thanks{Email: \protect\href{mailto:axelaraneda@mail.muni.cz}{\texttt{axelaraneda@mail.muni.cz}}}}
\maketitle

\lyxaddress{\begin{center}
\vspace{-2em} Institute of Financial Complex Systems \\ Department
of Finance\\ Masaryk University\\ 602 00 Brno, Czech Republic.
\par\end{center}}

\begin{center}
\vspace{-1em} This version: \today \vspace{1.5em}
\par\end{center}
\begin{abstract}
Fractional Brownian motion has become a standard tool to address long-range
dependence in financial time series. However, a constant memory parameter
is too restrictive to address different market conditions. Here we
model the price fluctuations using a multifractional Brownian motion
assuming that the Hurst exponent is a time-deterministic function.
Through the multifractional Ito calculus, both the related transition
density function and the analytical European Call option pricing formula
are obtained. The empirical performance of the multifractional Black-Scholes
model is tested by calibration of option market quotes for the SPX
index and offers best fit than its  counterparts based on standard
and fractional Brownian motions.


\vspace{2em}
\end{abstract}

\section{Introduction}

Since the Black and Scholes seminal paper \citep{black1973pricing},
diffusion processes driven by standard Brownian motions have been
consolidatedas the cornerstone of financial engineering, where one
of these features relies on the independence of the logarithmic returns.
This idea is also consistent with the efficient market hypothesis
(EMH) of Fama \citep{fama1970efficient}, another fundamental postulate
of modern finance, which implies, in its weak form, the absence of
a profitable strategy based on past information. However, even in
the early 60\textquoteright s B. Mandelbrot \citep{MANDELBROT1963}
challenged the idea of return independence, being established along
the years the long-range dependence (a.k.a. long-memory or power-like
decay in the returns autocorrelation) as a \textquoteleft stylized
fact\textquoteright{} in the analysis of financial time series \citep{Greene1977,Peters1989,Mandelbrot2013,Garzarelli2014,willinger1999stock,caporale2019long}.
Since these findings are in direct contrast to the EMH, the term \textquoteright fractal
market hypothesis\textquoteright{} was coined \citep{Peters1994}.

To address the memory effect, some researchers replaced the standard
Brownian motion driving the stochastic differential equation of the
price fluctuations with a suitable stochastic process. Certainly,
the most used is the fractional Brownian motion (fBm) \citep{necula2002option,araneda2020fractional,he2021fractional,costabile2023lattice}
which considers persistence by a power-law covariance structure.
Alternatives based on other anomalous diffusion processes, such as
the sub-fractional Brownian motion \citep{xu2019pricing,araneda2021sub,bian2021fuzzy,wang2022pricing},
as well as the use of fractional-order derivatives \citep{Golbabai2019,Golbabai2020,Nikan}
or fractional integrated econometric models \citep{Granger1980},
have been developed.

However, the assumption of a constant Holder regularity (Hurst exponent)
in financial time-series seems to be too rigid to address some particularities
of a market beyond tranquil periods, namely bull or bear markets,
and both memory and memory-less can be present in the same financial
data \citep{bianchi2005pathwise,bianchi2013modeling}. Indeed, some
scholars empirically state a time-varying behavior for the memory
parameter \citep{guangxi2014time,guedes2022efficiency}. In terms
of modelling, the mathematical tool compatible with this behavior
is called multifractional Brownian motion (mBm) \citep{peltier1995multifractional,benassi1997elliptic}.
This centered Gaussian process acts as a generalization of fBm in
the sense that it allows to the Hurst exponent becomes a time-deterministic\footnote{Some developments \citep{ayache2005multifractional,ayache2022moving}
extend this local behavior to non-deterministic cases or stochastic
processes.} local quantity. The implications of using mBm as the driven process
in price fluctuations are listed in ref. \citep{Bianchi2015}, and
among them is the compatibility with Lo's adaptive market hypothesis
\citep{lo2005reconciling} which dismiss the efficiency/inefficiency
dichotomy arguing that the level of efficiency changes on time around
the efficient state.

The literature offers some examples of the uses of mBm in option pricing.
For instance, Wang \citep{wang2010scaling} addresses the problem
under transaction cost, where a discrete-time setting obtains the
minimal value for a European Call by delta hedging arguments. In addition,
Mattera and Sciorio \citep{mattera2021option} elaborated a numerical
procedure to value a European Call option in a multifractional environment,
considering an autoregressive behavior for the Hurst exponent. On
the other hand, Corlay et al. \citep{corlay2014multifractional} arise
a multifractional version for both Hull \& White and log-normal SABR
stochastic volatility models with the aim to fit the shape of the
smile at different maturities. Similarly, Ayache and Peng \citep{ayache2012stochastic}
discuss parameter estimation for the integrated volatility driven
by mBm.

Our insight here is slightly different and focused on the analytical
results for option pricing in a continuous-time setting. First, we
assume that the noise behavior of the asset dynamics can be modeled
using an mBm with Hurst exponent described by a time-deterministic
function, and second, taking borrow some results based on stochastic
calculus related to mBm, the respective effective Fokker-Planck equation
is derived and solved, and consequently, the option pricing formula
is addressed.

The paper is organized as follows. First, we list some general properties
and auxiliary results for the fBm and mBm, particularly the multifractional
It\^o's lemma and the obtention of the related Fokker- Planck equation.
Later, we deal with the pricing procedure, focusing on an analytic
solution for the transition density and the proper pricing formula
using the actuarial approach. Next, an actual financial data experiment
shows the superior performance of the proposed approach compared to
the standard and fractional Black-Scholes formulas. Finally, the main
conclusions are listed 

\section{On the Multifractional Brownian motion}

A (normalized) fBm $B_{t}^{H}$ is a centered Gaussian process fully
determined by the following covariance function, with $t,s\geq0$
\citep{Mandelbrot1968}:

\begin{equation}
\mathbb{E}\left[B_{t}^{H}\cdot B_{s}^{H}\right]=\frac{1}{2}\left\{ \left|t\right|^{2H}+\left|s\right|^{2H}-\left|t-s\right|^{2H}\right\} \label{eq:covFBM}
\end{equation}

\noindent where $0<H<1$ is a constant parameter called the Hurst
exponent. And then, the second moment is given by:

\[
\text{var}\left(B_{t}^{H}\right)=\mathbb{E}\left[\left(B_{t}^{H}\right)^{2}\right]=t^{2H}
\]

It can be shown that, according to the value of $H$, the increments
are i) positive correlated with long-range dependence for $H>1/2$,
ii) negatively correlated and short range dependent for $H<1/2$,
or iii) independent if $H=1/2$. In the latter the process matches
to the standard Brownian motion $\left(B_{t}^{H=1/2}=B_{t}\right)$. 

The main features for the fBm include self-similarity and sationary
increments, as in the case of Brownian motion. However, for $H\neq1/2,$
the process is not Markov and nor a semi-martingale. It means that
the standard Ito calculus is not suitable and the fractional Ito lemma
should be considered instead \citep{bender2003ito}.

On the other hand, the mBm is also a centered Gaussian proceess, which
extend the fBm by the way of a time-dependent Hurst exponent (H\"{o}lderian
funcion), behaving locally as a fBm \citep{peltier1995multifractional}.
An standard mBm $W_{h\left(t\right)}$ is a centered Gaussian process
formally defined by its covariance \citep{Ayache2000}:

\begin{equation}
\text{\ensuremath{\mathbb{E}\left[W_{h\left(t\right)}\cdot W_{h\left(s\right)}\right]=D\left(t,s\right)\left[t^{h(t)+h(s)}+s^{h(t)+h(s)}+\left|t-s\right|^{h(t)+h(s}\right]}}\label{eq:covmBm}
\end{equation}
\noindent where

\[
D\left(t,s\right)=\frac{\sqrt{\Gamma\left(2H_{t}+1\right)\Gamma\left(2H_{s}+1\right)\sin\left(\pi H_{t}\right)\sin\left(\pi H_{s}\right)}}{2\Gamma\left(H_{t}+H_{s}+1\right)\sin\left[\frac{\pi}{2}\left(H_{t}+H_{s}\right)\right]}
\]

\noindent and $h:\left[0,\infty\right)\rightarrow\left[l,m\right]\subset\left(0,1\right)$.

As pointed by Ayache et al. \citep{Ayache2000}, the covariance structure
(\ref{eq:covmBm}) exhibits long-range dependence. Moreover, the variance
is expressed as:

\[
\mathbb{E}\left[\left(W_{h\left(t\right)}\right)^{2}\right]=t^{2h\left(t\right)}
\]

Even though the mBm is a generalization for the fbm, the former is
not self-similar and doesn't have stationary increments. It also lacks
of the Markov poperty and semimartingality. Thus, as in the case of
the fBm, an stochastic calculus with respect to multifractional Brownian
processes has been developed \citep{lebovits2014white}, and summarized
at next.

\paragraph{}

 Let $F\in C^{2}\left(\mathbb{R}\right)$ and $y_{t}$ a generic stochastic
process driven by a multifractional Brownian motion:

\begin{equation}
\text{d}y_{t}=a\left(y_{t},t\right)\text{dt}+b\left(t,y_{t}\right)\text{d}W_{h\left(t\right)}\label{eq:gen}
\end{equation}

Then, the following equality holds:

\begin{eqnarray}
\text{d}F\left(t,y_{t}\right) & = & \frac{\partial F}{\partial t}\text{d}t+\frac{\partial F}{\partial y_{t}}\text{d}y_{t}+\frac{1}{2}\left\{ \frac{\text{d}}{\text{d}t}\left[t^{2h\left(t\right)}\right]\right\} b^{2}\frac{\partial^{2}F}{\partial t^{2}}\text{d}t\nonumber \\
 & = & \left\{ \frac{\partial F}{\partial t}+a\frac{\partial F}{\partial y_{t}}+b^{2}t^{2h(t)-1}\left[h'\left(t\right)t\ln t+h\left(t\right)\right]\frac{\partial^{2}F}{\partial t^{2}}\right\} \text{d}t+b\frac{\partial F}{\partial y_{t}}\text{d}W_{h\left(t\right)}\label{eq:mfI}
\end{eqnarray}

For a constant function $h(t)=H,$ the above theorem is reduced to
the Fractional Ito formula addressed by Bender \citep{bender2003ito},
while for the fixed value $h(t)=H=1/2$, the standard It\^o's lemma
is recovered.

With the help of the multifractional Ito lemma (\ref{eq:mfI}), one
can derive the so called ``Effective Fokker-Planck Equation'' (EFPE)
for the stochastic process $y_{t}$ ruled by the SDE (\ref{eq:gen}).
Let considers a twice-differentiable scalar function $g\left(y_{t}\right).$

\paragraph{}

 Using the multifractional It\^o calculus and taking expectations,
we get:

\[
\frac{\text{d}\mathbb{E}\left(g\right)}{\text{d}t}=\mathbb{E}\left(a\frac{\partial g}{\partial y}\right)+\mathbb{E}\left\{ b^{2}t^{2h(t)-1}\left[h'\left(t\right)t\ln t+h\left(t\right)\right]\frac{\partial^{2}g}{\partial y^{2}}\right\} 
\]

Recalling the definition of expectations by means of the transition
density function $P$, and after some calculus, the effective Fokker-Planck
related to the process (\ref{eq:gen}) emerges:

\begin{equation}
\frac{\partial P}{\partial t}=t^{2h(t)-1}\left[h'\left(t\right)t\ln t+h\left(t\right)\right]\frac{\partial^{2}\left(b^{2}P\right)}{\partial y^{2}}-\frac{\partial\left(aP\right)}{\partial y}\label{eq:TDF}
\end{equation}

The above equation matches to Gaussian diffusion with drift a per
unit of time and an ``effective'' variance $b^{2}\int t^{2h(t)-1}\left[h'\left(t\right)t\ln t+h\left(t\right)\right]\text{d}t=b^{2}t^{2h(t)}$.

\section{The multifractional Black-Scholes model and its transition density
function}

Let's start with the standard geometric Brownian motion, Under the
real-world physical measure $\mathbb{P}$ it obeys:

\[
\text{d}S_{t}=\text{\ensuremath{\mu S_{t}\text{d}t}+\ensuremath{\sigma S_{t}W_{t}}}
\]

\noindent where the constant values $\mu$ and $\sigma$ represent
the yearly drift and volatility for the instantaneous return, and
$W_{t}$ an standard Gauss-Wiener process. Now in order to equipped
it with a time-varying long memory, we replace $W_{t}$ by a multifractional
Brownian motion $W_{t\left(h\right)}$:

\[
\text{d}S_{t}=\text{\ensuremath{\mu S_{t}\text{d}t}+\ensuremath{\sigma S_{t}W_{h\left(t\right)}}}
\]

The Holderian function of the mBm $W_{t(h)}$; i.e., $H(t)$, is assumed
known and time-deterministic (see for example ref. \citep{corlay2014multifractional}
for the case of a time-dependent sinusoidal function).

By the substitution $x_{t}=\ln S_{t}-\mu t$ , the multifractional
It\^o's lemma (Eq. \ref{eq:mfI}) leads to:

\begin{equation}
\text{d}x_{t}=-\sigma^{2}t^{2h(t)-1}\left[h'\left(t\right)t\ln t+h\left(t\right)\right]\text{d}t+\sigma W_{h\left(t\right)}\label{eq:dx}
\end{equation}

According to the multifractional Fokker-Planck equation, setting $a=-\sigma^{2}t^{2h(t)-1}\left[h'\left(t\right)t\ln t+h\left(t\right)\right]$
and $b=\sigma$ in Eq. \ref{eq:TDF}, the transition density $P=P\left(x_{t},t\right)$
related to the process (\ref{eq:dx}) obeys:

\begin{equation}
\frac{\partial P}{\partial t}=\sigma^{2}t^{2h(t)-1}\left[h'\left(t\right)t\ln t+h\left(t\right)\right]\left[\frac{\partial P}{\partial x}+\frac{\partial^{2}P}{\partial x^{2}}\right]\label{eq:FP_x}
\end{equation}

Using the time substitution:

\[
\bar{t}=\sigma^{2}t^{2h\left(t\right)}
\]

\noindent and defining the moving frame of reference:

\[
\bar{x}=x+\frac{\bar{t}}{2}
\]

Eq. (\ref{eq:FP_x}) goes to:

\[
\frac{\partial P}{\partial\bar{t}}=\frac{1}{2}\frac{\partial^{2}P}{\partial\bar{x}^{2}}
\]

The fundamental solution for the above equation (heat kernel with
constant thermal difussivity equal to $1/2$) is well known and given
by:

\begin{equation}
P\left(\bar{x},\bar{t}\right)=\frac{1}{\sqrt{2\pi\bar{t}}}\exp\left[-\frac{\left(\bar{x}-\bar{x}_{0}\right)}{2\bar{t}}\right]\label{eq:Pxt}
\end{equation}

\noindent where $P\left(\bar{x},0\right)=P\left(\bar{x}_{0}\right)=\delta\left(\bar{x}_{0}\right)$.
The initial condition is given by knowing the state of the asset at
the inception time; i.e., $S\left(t=0\right)=S_{0}=\text{e}^{x_{0}}=\text{e}^{\bar{x}_{0}}$.

In fact, according to Eq. (\ref{eq:Pxt}), $P\left(\bar{x},\bar{t}\right)$
describes a Gaussian probability density function centered at $\bar{x}_{0}$
(expected value) and variance $\bar{t}$. 

Coming back to the variable $x$ and the original time $t$, the transition
density is expressed as:

\[
P\left(x,t\right)=\frac{1}{\sqrt{2\pi\sigma^{2}t^{2h\left(t\right)}}}\exp\left[-\frac{\left(x-x_{0}+\frac{1}{2}\sigma^{2}t^{2h\left(t\right)}\right)^{2}}{2\sigma^{2}t^{2h\left(t\right)}}\right]
\]

From the previous result, we can compute the first moment for asset
price in a future time $t=T$ subject to its value at the inception
$t=0$:

\begin{eqnarray}
\mathbb{E^{P}}\left(S_{T}\right) & = & \int_{0}^{\infty}S_{T}P\left(S_{T},T\right)\text{d}S_{T}\nonumber \\
 & = & \int_{0}^{\infty}\text{e}^{x_{T}+\mu T}P\left(x_{T},T\right)\text{d}x_{T}\nonumber \\
 & = & \frac{\text{e}^{x_{0}+\mu T}}{\sqrt{2\pi\sigma^{2}T^{2h\left(T\right)}}}\int_{0}^{\infty}\exp\left[-\frac{\left(x-x_{0}-\frac{1}{2}\sigma^{2}T^{2h\left(T\right)}\right)^{2}}{2\sigma^{2}T^{2h\left(T\right)}}\right]\text{d}x_{T}\nonumber \\
 & = & \frac{S_{0}\text{e}^{\mu T}}{\sqrt{2\pi}}\int_{0}^{\infty}\exp\left[-\frac{u^{2}}{2}\right]\text{d}u\nonumber \\
 & = & S_{0}\text{e}^{\mu T}\label{eq:S0}
\end{eqnarray}

\noindent where no differences appear concerning the expectation
in the classical Black Scholes world, while in the second central
moment, they differ\footnote{In the standard Geometric Brownian motion, the variance for the price
at time $T$ is equal to $S_{0}\text{e}^{2\mu T}\left(\text{e}^{\sigma^{2}T}-1\right)$.}:

\begin{eqnarray}
\mathbb{E^{P}}\left[\left(S_{T}-\mathbb{E^{P}}\left(S_{T}\right)\right)^{2}\right] & = & \int_{0}^{\infty}S_{T}^{2}P\left(S_{T},T\right)\text{d}S_{T}-S_{0}\text{e}^{2\mu T}\nonumber \\
 & = & \int_{0}^{\infty}\text{e}^{2x_{T}+2\mu T}P\left(x_{T},T\right)\text{d}x_{T}-S_{0}\text{e}^{2\mu T}\nonumber \\
 & = & \frac{\text{e}^{x_{0}+\mu T}}{\sqrt{2\pi\sigma^{2}T^{2h\left(T\right)}}}\int_{0}^{\infty}\exp\left[-\frac{\left(x-x_{0}-\frac{1}{2}\sigma^{2}T^{2h\left(T\right)}\right)^{2}}{2\sigma^{2}T^{2h\left(T\right)}}\right]\text{d}x_{T}-S_{0}\text{e}^{2\mu T}\nonumber \\
 & = & \frac{S_{0}\text{e}^{\mu T}}{\sqrt{2\pi}}\int_{0}^{\infty}\exp\left[-\frac{u^{2}}{2}\right]\text{d}u-S_{0}\text{e}^{2\mu T}\nonumber \\
 & = & S_{0}\text{e}^{2\mu T}\left(\text{e}^{\sigma^{2}T^{2h\left(T\right)}}-1\right)\label{eq:S0-1}
\end{eqnarray}

\section{The option pricing formula}

Under mBm diffusion there is no equivalent martingale measure, so
the standard risk risk-neutral pricing is not available \citep{mattera2021option}.
However, we can apply the actuarial approach \citep{bladt1998actuarial}
in order to get an option pricing formula without the semi-martingale
assumption. The idea of this alternative method is to get the derivative
valuation by insurance considerations using the fair premium principle
under the physical probability measure. The fair option premium is
computed by expectations of the present value issuer's loss. The main
advantage is the lack of any economic assumption, such as arbitrage-free
or market completeness, without using the equivalent martingale (risk-neutral)
measure. Recent applications of the actuarial approach include the
valuation of currency \citep{Falkowski2022} and vulnerable options
\citep{Cheng2023} in a fractional Brownian motion setting.

Let $\text{e}^{\mu T}=\frac{\mathbb{E^{P}}\left(S_{T}\right)}{S_{0}}$
the expected rate of return for the asset $S$ at time $T$ (see Eq.
\ref{eq:S0}). By the actuarial approach, the fair premium for a vanilla
European Call option with maturity $T$ and exercise price $K$ is
given by \citep{bladt1998actuarial}:

\[
C\left(K,T\right)=\mathbb{E^{P}}\left[\left(\text{e}^{-\mu T}S_{T}-\text{e}^{-rT}K\right)^{+}\right]
\]

Since:

\begin{eqnarray*}
\text{e}^{-\mu T}S_{T}>\text{e}^{-rT}K & \iff & \text{e}^{x_{T}}>\text{e}^{-rT}K\\
 & \iff & x_{T}>\ln K-rT
\end{eqnarray*}

\noindent we have:

\begin{eqnarray}
C\left(K,T\right) & = & \int_{\ln K-rT}^{\infty}\left(\text{e}^{x_{T}}-\text{e}^{-rT}K\right)P\left(x_{T},T\right)\text{d}x_{T}\nonumber \\
 & = & \int_{\ln K-rT}^{\infty}\text{e}^{x_{T}}P\left(x_{T},T\right)\text{d}x_{T}-K\text{e}^{-rT}\int_{\ln K-rT}^{\infty}P\left(x_{T},T\right)\text{d}x_{T}\label{eq:C}
\end{eqnarray}

Given that,

\begin{eqnarray*}
\int_{\ln K-rT}^{\infty}\text{e}^{x_{T}}P\left(x_{T},T\right)\text{d}x_{T} & = & \frac{\text{e}^{x_{T}}}{\sqrt{2\pi\sigma^{2}T^{2h\left(t\right)}}}\exp\left[-\frac{\left(x-x_{0}+\frac{1}{2}\sigma^{2}T^{2h\left(T\right)}\right)^{2}}{2\sigma^{2}T^{2h\left(T\right)}}\right]\\
 & = & \frac{\text{e}^{x_{0}}}{\sqrt{2\pi\sigma^{2}T^{2h\left(T\right)}}}\int_{0}^{\infty}\exp\left[-\frac{\left(x-x_{0}-\frac{1}{2}\sigma^{2}T^{2h\left(T\right)}\right)^{2}}{2\sigma^{2}T^{2h\left(T\right)}}\right]\text{d}x_{T}\\
 & = & -\frac{\text{e}^{x_{0}}}{\sqrt{2\pi}}\int_{\frac{x_{0}-\ln K+rT+\frac{1}{2}\sigma^{2}T^{2h\left(T\right)}}{\sqrt{\sigma^{2}T^{2h\left(T\right)}}}}^{\infty}\text{e}^{-\frac{v^{2}}{2}}\text{d}v\\
 & = & \text{e}^{x_{0}}N\left(d_{1}^{h(t)}\right)
\end{eqnarray*}

\begin{eqnarray*}
\int_{\ln K-rT}^{\infty}P\left(x_{T},T\right)\text{d}x_{T} & = & \frac{1}{\sqrt{2\pi\sigma^{2}T^{2h\left(t\right)}}}\int_{\ln K-\mu T}^{\infty}\exp\left[-\frac{\left(x-x_{0}+\frac{1}{2}\sigma^{2}T^{2h\left(t\right)}\right)^{2}}{2\sigma^{2}T^{2h\left(t\right)}}\right]\text{d}x_{T}\\
 & = & -\frac{1}{\sqrt{2\pi}}\int_{\frac{x_{0}-\ln K+rT-\frac{1}{2}\sigma^{2}T^{2h\left(T\right)}}{\sqrt{\sigma^{2}T^{2h\left(T\right)}}}}^{\infty}\text{e}^{-\frac{w^{2}}{2}}\text{d}w\\
 & = & N\left(d_{2}^{h(t)}\right)
\end{eqnarray*}

\noindent where $N\left(\cdot\right)$ stands for the standard normal
cumulative density and:

\[
d_{1}^{h(t)}=\frac{x_{0}-\ln K+rT+\frac{1}{2}\sigma^{2}T^{2h\left(T\right)}}{\sqrt{\sigma^{2}T^{2h\left(T\right)}}}=\frac{\ln\left(\nicefrac{S_{0}}{K}\right)+rT+\frac{1}{2}\sigma^{2}T^{2h\left(T\right)}}{\sqrt{\sigma^{2}T^{2h\left(T\right)}}}
\]

\[
d_{2}^{h(t)}=d_{1}^{h(t)}-\sqrt{\sigma^{2}T^{2h\left(T\right)}}
\]

Consequently, after replace the above computations into Eq. (\ref{eq:C}),
we can arrive at the pricing for a European Call under multifractional
diffusion:

\begin{equation}
C\left(K,T\right)=S_{0}N\left(d_{1}^{h(t)}\right)-K\text{e}^{-rT}N\left(d_{2}^{h(t)}\right)\label{eq:BS}
\end{equation}

The formula (\ref{eq:BS}) is also a generalization of the previous
approaches. If $h(t)=H$ is fixed to some value in its domain, the
above result is equivalent to the fractional Black-Scholes formula
\citep{necula2002option}, while for $h\left(t\right)=1/2$, the standard
Black-Scholes premium is recovered. The main difference among them
resides on the volatility side. While in the classical Black-Scholes
model the variance scale linear on time, in the fractional approach
does it by the constant power-law t\textasciicircum\{2H\} In the
multifractional setting, the power scaling keeps, but now through
by a time-dependent exponent.

\section{Numerical results}

To evaluate the empirical performance of the proposed model, we assess
its ability to reproduce actual option market quotes, comparing it
with its fractional and classical counterparts. Since models \emph{\`a la}
Black-Scholes return only one option price for a given maturity, we
consider only a fixed strike over different maturities. Consequently,
we will rank the models according to their capability to catch the
option price term structure.

We select Call option prices for the S\&P 500 index (SPX) quoted on
March 24, 2023 (closing prices) to proceed. We consider only at-the-money
options (stock closing price=\$3970.99, strike price=3970) and a range
of maturities from 1 day to 5 months. As a proxy for the risk-free
rate-of-interest (r), we use the 13-week T-Bill rate on the inception
time (4.5013\%). 

\begin{figure}
\includegraphics[width=1\textwidth]{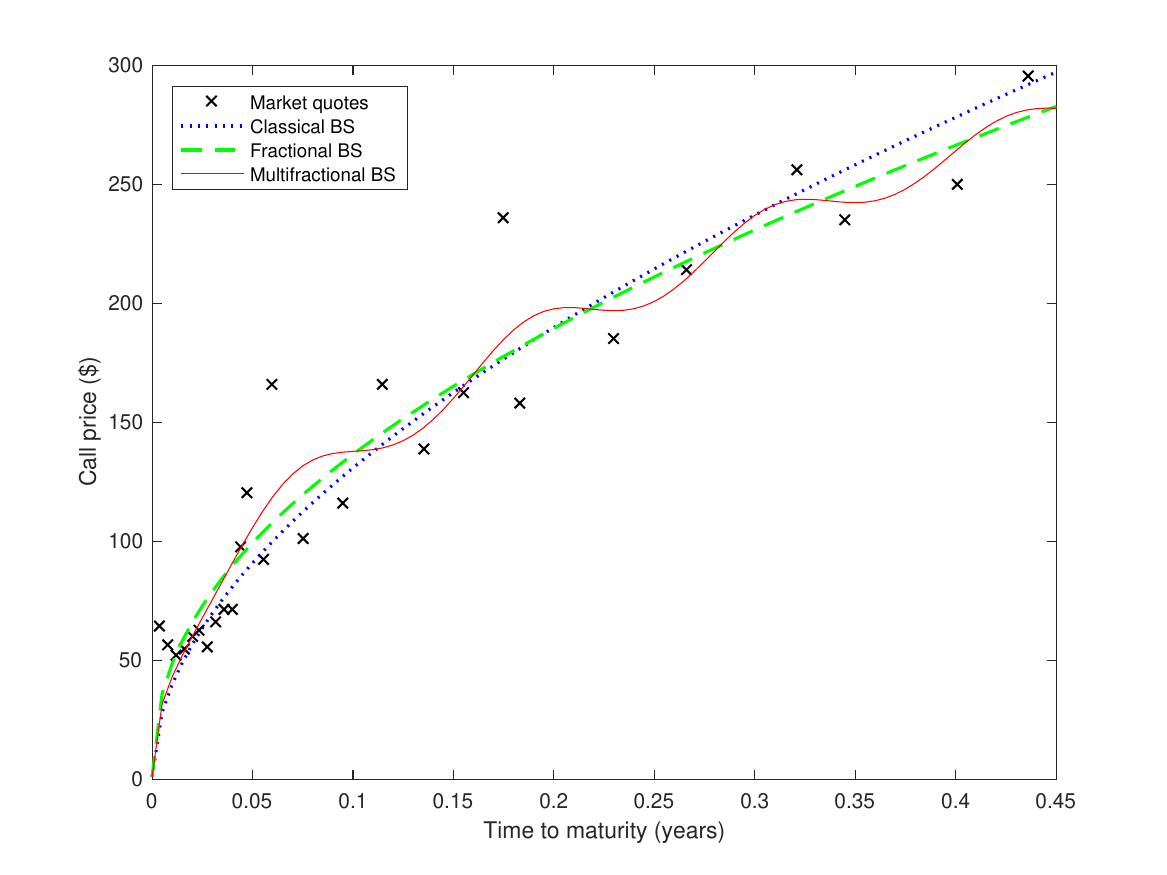}

\caption{ATM Call option quotes and the empirical fit of the pricing models\label{fig:Performance-of-the}}
\end{figure}

To go forward with the calibration we use the 252 yearly days convention.
For the multifractional Black-Scholes, as in ref. \citep{corlay2014multifractional},
we select a 6-week ($\sim30$ trading days) periodic sinusoidal function
for the point-wise regularity, particularly, $\hat{h}(t)=A\cos\left[2\pi\left(\frac{252}{30}\right)t+B\right]+C$;
where $A$, $B$, and $C$, in addition to $\sigma$, are parameters
that should be estimated.The optimal set of parameters is obtained
by minimizing the squared residuals between the model output and actual
data, using lsqnonlin function in Matlab. Fig. \ref{fig:Performance-of-the}
shows the ATM Call option prices as a function of the maturity jointly
with the prices returned by the standard, fractional, and multifractional
Black-Scholes formulas. The flexibility given by the function form
of $H(t)$ leads to superior performance for the multifractional BS
model, where the mean square errors of the market quotes compared
with model prices are lower (456.8) than the fractional (493.7), and
the standard BS (555.5).

\section{Summary}

We have modeled the price fluctuation by means of a Geometric Brownian
motion driven by a multifractional Brownian motion where the Hurst
exponent is an exclusive function of time. Our main result here is
the obtention of the analytical multifractional Black-Scholes formula
by means of the multifractional Ito calculus, the related Fokker-Planck
equation, and the actuarial approach to price option under the physical
measure $\mathbb{P}$. Since mBm is a generalization for both Bm and
fBm, the classical and fractional Black Scholes option pricing are
recovered. On the experimental side, a numerical validation with real
market data was carried out using SPX ATM European Call options quotes,
exhibing higher  empirical fit if we consider a time-dependent Hurst
exponent. The option pricing under several functions for the Hurst
exponent and different extensions of the multifractional Brownian
motion are field of further research.

\bibliographystyle{28_Users_axelaraneda_Desktop_Research_Hurst_ws-fnl}
\bibliography{27_Users_axelaraneda_Desktop_Research_Hurst_multifractional}

\end{document}